\documentstyle[aps,epsfig]{revtex}
\def \blankline{\vspace{0.4 cm}}

\begin {document}
\draft

\title {\bf Muon `Depth -- Intensity' Relation Measured by LVD 
Underground Experiment and Cosmic-Ray Muon Spectrum at Sea Level}
\blankline
\author{
{\bf The LVD Collaboration}\\
\vspace{0.3cm}
M.Aglietta$^{16}$, B.Alpat$^{13}$, E.D.Alyea$^{7}$, P.Antonioli$^{1}$,
G.Badino$^{16}$, G.Bari$^{1}$, M.Basile$^{1}$,
V.S.Berezinsky$^{10}$, F.Bersani$^{1}$, M.Bertaina$^{16}$,
R.Bertoni$^{16}$, G.Bonoli$^{1}$, A.Bosco$^{2}$,
G.Bruni$^{1}$, G.Cara Romeo$^{1}$,
C.Castagnoli$^{16}$, A.Castellina$^{16}$, A.Chiavassa$^{16}$,
J.A.Chinellato$^{3}$, 
L.Cifarelli$^{1,\dagger}$, 
F.Cindolo$^{1}$,
G.Conforto$^{17}$,
A.Contin$^{1}$, V.L.Dadykin$^{10}$,
A.De Silva$^{2}$, M.Deutsch$^{8}$, P.Dominici$^{17}$,
L.G.Dos Santos$^{3}$, L.Emaldi$^{1}$,
R.I.Enikeev$^{10}$, F.L.Fabbri$^{4}$, W.Fulgione$^{16}$,
P.Galeotti$^{16}$, C.Ghetti$^{1}$,
P.Ghia$^{16}$, P.Giusti$^{1}$, R.Granella$^{16}$, F.Grianti$^{1}$,
G.Guidi$^{17}$,
E.S.Hafen$^{8}$, P.Haridas$^{8}$, G.Iacobucci$^{1}$, N.Inoue$^{14}$,
E.Kemp$^{3}$, F.F.Khalchukov$^{10}$, E.V.Korolkova$^{10}$,
P.V.Korchaguin$^{10}$, V.B.Korchaguin$^{10}$, V.A.Kudryavtsev$^{10,\ddagger}$,
K.Lau$^{6}$, M.Luvisetto$^{1}$, G.Maccarone$^{4}$,
A.S.Malguin$^{10}$, R.Mantovani$^{17}$,
T.Massam$^{1}$,
B.Mayes$^{6}$, A.Megna$^{17}$, C.Melagrana$^{16}$,
N.Mengotti Silva$^{3}$,
C.Morello$^{16}$, J.Moromisato$^{9}$,
R.Nania$^{1}$,
G.Navarra$^{16}$, L.Panaro$^{16}$,
L.Periale$^{16}$, A.Pesci$^{1}$, P.Picchi$^{16}$, L.Pinsky$^{6}$,
I.A.Pless$^{8}$, J.Pyrlik$^{6}$, V.G.Ryasny$^{10}$,
O.G.Ryazhskaya$^{10}$, O.Saavedra$^{16}$, K.Saitoh$^{15}$,
S.Santini$^{17}$, 
G.Sartorelli$^{1}$, 
M.Selvi$^{1}$, 
N.Taborgna$^{5}$,
V.P.Talochkin$^{10}$,
J.Tang$^{8}$, G.C.Trinchero$^{16}$, S.Tsuji$^{11}$, A.Turtelli$^{3}$,
I.Uman$^{13}$, P.Vallania$^{16}$, G. Van Buren $^{8}$,
S.Vernetto$^{16}$,
F.Vetrano$^{17}$, C.Vigorito$^{16}$, E. von Goeler$^{9}$,
L.Votano$^{4}$, T.Wada$^{11}$,
R.Weinstein$^{6}$, M.Widgoff$^{2}$,
V.F.Yakushev$^{10}$, I.Yamamoto$^{12}$,
G.T.Zatsepin$^{10}$, A.Zichichi$^{1}$
}
\vspace{0.3cm}
\address{
\noindent $^{1}${\it University of Bologna and INFN-Bologna, Italy}\\
$\,^{2}${\it Brown University, Providence, USA}\\
$\,^{3}${\it University of Campinas, Campinas, Brazil}\\
$\,^{4}${\it INFN-LNF, Frascati, Italy}\\
$\,^{5}${\it INFN-LNGS, Assergi, Italy}\\
$\,^{6}${\it University of Houston, Houston, USA}\\
$\,^{7}${\it Indiana University, Bloomington, USA}\\
$\,^{8}${\it Massachusetts Institute of Technology, Cambridge, USA}\\
$\,^{9}${\it Northeastern University, Boston, USA}\\
$^{10}${\it Institute for Nuclear Research, Russian Academy of
Sciences, Moscow, Russia}\\
$^{11}${\it Okayama University, Okayama, Japan}\\
$^{12}${\it Okayama University of Science, Okayama, Japan}\\
$^{13}${\it University of Perugia and INFN-Perugia, Italy}\\
$^{14}${\it Saitama University, Saitama, Japan}\\
$^{15}${\it Ashikaga Institute of Technology, Ashikaga, Japan}\\
$^{16}${\it Institute of Cosmo-Geophysics, CNR, Torino, University
of Torino and} \\
\indent {\it INFN-Torino, Italy}\\
\noindent $^{17}${\it University of Urbino and INFN-Firenze, Italy}\\
\vspace{0.3cm}
$^{\dagger}${\it now at University of Salerno and INFN-Salerno, Italy}\\ 
$^{\ddagger}${\it now at University of Sheffield, United Kingdom}
}
\maketitle

\begin{abstract}
We present the analysis of the muon events with all muon multiplicities
collected during 21804
hours of operation of the first LVD tower. The measured angular
distribution of muon intensity has been converted to the
`depth -- vertical intensity' relation in the depth range from 3
to 12 km w.e.. The analysis of this relation allowed to derive
the power index, $\gamma$, of the primary all-nucleon
spectrum: $\gamma=2.78 \pm 0.05$. The `depth -- vertical intensity' 
relation has been converted to standard rock and the comparison
with the data of other experiments has been done.
We present also the derived vertical muon spectrum at sea level.
\end{abstract}

\section { Introduction }

During the last 30 years the cosmic-ray muon energy spectrum
has been studied in many
experiments using different methods. These methods can be combined
into 3
groups: i) the direct measurements of the muon energy spectrum at
the sea
level using magnetic spectrometers \cite{MARS,DEIS,MUTRON}
(the spectrum
up to about 10 TeV was measured at the zenith angles near horizon);
ii)
the measurements of the energy spectrum of cascades produced by muons
at
shallow depth \cite{ASD,Baksans,MIPhI,MSU}; iii) the measurement of the
depth--intensity curve deep underground
\cite{KGF,Baksan,NUSEX,MACROc,LVD,Frejusf}.

Since the spectrum of primary cosmic-ray nucleons has the
power-law form
with the power index $\gamma$, the spectrum of $\pi-$ and $K-$ mesons
produced by primaries should have also the power-law form with the power
index $\gamma_{\pi,K}$. If the scaling hypothesis is valid in the
fragmentation region at high energies, the value of $\gamma_{\pi,K}$ is
approximately equal to $\gamma$. The muon spectrum at the sea level
has a more complex
form due to the competition between the interaction and decay of their
parents. Moreover, the flux of muons produced by pions and kaons
strongly
depends on the zenith angle because the interaction path length of
mesons
in the atmosphere varies with the zenith angle $\theta$. The muon
spectrum
follows the power-law dependence only at high energies
$E_{\mu}>>E_{\pi}^{cr},E_K^{cr}$, where $E_{\pi}^{cr}$ and
$E_K^{cr}$ are
the critical energies of pions and kaons in the atmosphere.
For $\theta=0$
(vertical) this condition is quite satisfied at $E_{\mu}>$ 2 TeV.
The power
index of the muon spectrum, $\gamma_{\mu}$, in this case is more
by 1 than
the value of $\gamma_{\pi,K}$.

Despite the numerous experiments which studied the muon
spectrum, there are
the discrepancies in the published results (values of $\gamma$,
$\gamma_{\pi,K}$ or $\gamma_{\mu}$). Most of the experiments carried out
in the last 20 years gave the values of $\gamma$ (or $\gamma_{\pi,K}$)
in the range 2.60 -- 2.80. 
But the dispersion of the results is greater than the
statistical and systematic errors published by the authors. Even the
values of $\gamma$, obtained using one method (for example, by
the measurement of the depth-intensity curve) have a large dispersion.
Such
discrepancy can be due to either the difference between the data
themselves,
or between the calculated muon intensities used to fit the data,
or both.

The muon intensities deep underground have been calculated
by many authors
using different methods (Monte-Carlo simulation of the muon transport,
numerical solution of the kinetic equations etc., see, for example,
\cite
{Gurentsova,Gurentsov,Castellina,Bilokon,Lipari,Kudryavtsev,Naumov,Lagutin}).
However, the discrepancy between the muon survival probabilities and,
hence,
between the muon intensities, calculated by different authors,
is quite large
and can result in the significant discrepancy between the final
values of $\gamma$.

The LVD (Large Volume Detector), located underground,
can measure
the atmospheric muon intensities from 3000 hg/cm$^2$ to 12000 hg/cm$^2$
and above (which
correspond to the muon energies at the sea level from 1.5 TeV to
40 TeV) at the zenith angles from $0^o$ to $90^o$. This
allows us to study the muon spectrum and their characteristics at the
energies 1.5--40 TeV (which correspond to the energies of primaries
of about 10--400 TeV).

In a previous paper \cite{LVD} we have presented our
measurement of the muon depth--intensity curve and the evaluation of the
power index of the meson spectrum in the atmosphere using the
depth--intensity relation
for single muon events. The muon survival probabilities, used to
obtain the value of $\gamma_{\pi,K}$ in \cite{LVD}, have been presented
in \cite{LVDR}. They have been calculated using the muon interaction
cross-sections from \cite{BBb,BBn,KP}. After the publication of these
results, new calculation of the cross-section of muon bremsstrahlung
and of the corrections to the knock-on electron production cross-section
have been done \cite{KKP}. In the present analysis we have taken into
account the corrections proposed in \cite{KKP} and we have estimated
the uncertainties of $\gamma$ ($\gamma$ is assumed to be equal to
$\gamma_{\pi,K}$) due to the uncertainties of the
cross-sections used to simulate the muon transport through the rock.
The analysis is based on an increased statistics comparing
with the previous publications and refers to events with all
muon multiplicities.

In Section 2 the detector and the procedure of data processing
together with the conversion of the muon intensity to vertical
are briefly described. In Section 3
the results of the analysis of the `depth -- vertical muon intensity'
distribution ($I_{\mu}(x)$) are
presented. In Section 4 the `depth -- vertical intensity'
relation in standard rock is compared with the data of other
underground experiments. In Section 5 we present the derived
muon spectrum at sea level.
Section 6 contains the conclusions.

\section { LVD and data processing }

The LVD (Large Volume Detector) is located in the
underground Gran Sasso Laboratory at a minimal depth of about
3000 hg/cm$^2$.
The LVD will consist of 5 towers. The 1st tower is running since
June, 1992,
and the 2nd one - since June, 1994. The data presented here were
collected with the 1st LVD tower during 21804 hours of live time.

The 1st LVD tower contains 38 identical modules \cite{LVDa}.
Each module con\-sists of 8 scintillation counters and 4 layers of
limited
streamer tubes (tracking detector) attached to the bottom and to one
vertical side of the supporting structure. A detailed description of the
detector was given in \cite{LVDa}. One LVD tower has the dimensions of
$13 \times 6.3 \times 12$ m$^3$.

The LVD measures the atmospheric muon intensities from
3000 hg/cm$^2$ to more than 12000 hg/cm$^2$ (which
correspond to the median muon energies at the sea level from 1.5 TeV to
40 TeV) at the zenith angles from $0^o$ to $90^o$
(on the average, the larger depths correspond to higher zenith angles).

We have used in the analysis the muon events with all multiplicities,
as well as the sample of single muons. Our basic results have been
obtained with all muon sample. This sample contains about 2 millions of
reconstructed muon tracks.

The acceptances for each angular bin have been calculated using
the simulation of muons passing through LVD taking into account
muon interactions with the detector materials and the detector
response. The acceptances for both single and multiple muons were
assumed to be the same.

As a result of the data processing the angular distribution
of the number of detected muons
$N_{\mu}(\phi, \cos \theta)$ has been obtained. The angular bin
width $1^o \times 0.01$ has been used. The analysis refers to
the angular bins for which
the efficiency of the muon detection and track reconstruction is greater
than 0.03. We have excluded from the analysis the angular bins with
a large variation of depth.

The measured $N_{\mu}(\phi, \cos \theta)$-distribution has been
converted to the `depth -- vertical muon intensity' relation,
$I_{\mu}(x)$, using the formula:

\begin{equation}
I_{\mu}(x_m)=
{{\sum_{ij} N_{\mu}(x_m(\phi_j,\cos \theta_i))
\cdot \cos \theta_i^{\star\star}} \over
{\sum_{ij} (A(x_m(\phi_j,\cos \theta_i))
\epsilon(x_m(\phi_j,\cos \theta_i)) \cdot \Omega_{ij} \cdot T)}}
\label{depth-angle}
\end{equation}

where the summing up has been done over all angular bins 
$(\phi_j,\cos\theta_i)$
contributing to the depth $x_m$; $A(x_m(\phi_j,\cos \theta_i))$ is
the cross-section of the
detector in the plane perpendicular to the muon track at the angles
($\phi_j,\cos \theta_i$); $\epsilon(x_m(\phi_j,\cos \theta_i))$ is the
efficiency of muon detection and reconstruction; $\Omega_{ij}$ is the solid
angle for the angular bin; $T$ is the live time; and
$\cos \theta_i^{\star\star}=I_{\mu}^c(x_m,\cos \theta=1)/
I_{\mu}^c(x_m,\cos \theta_i)$ is the ratio of predicted muon
intensity at $\cos \theta=1$ to that at $\cos \theta_i$.
To obtain the values of $\cos \theta_i^{\star\star}$
the eq. (\ref{Gaisser spectrum}) (see below) with the 
previously estimated
parameters of the muon spectrum at sea level \cite{LVD,LVDR}
and the calculated survival probabilities
have been used. Actually, the values of $\cos \theta_i^{\star\star}$
do not depend much on the parameters of the muon spectrum at
sea level (normalization factor $A$ and power index $\gamma$) for
any reasonable values of $\gamma$.
The factor $\cos \theta_i^{\star\star}$
is different from the simple $\cos \theta$ - law used to convert the
muon intensities to vertical in \cite{LVD}.

In the calculations of $\cos \theta_i^{\star\star}$ the contribution
of the prompt muons from charmed particle decay has been neglected.
As it is shown in \cite{LVDP,propro} the ratio of prompt muons to pions
according to LVD data does not exceed $2 \cdot 10^{-3}$ at 95\%
confidence level. This means that at median depths (no more than 
6--7 km w.e.) the fraction of prompt muons with respect to conventional 
muons does not exceed 10\% at vertical. And there is also a little
probability that this fraction is more than 10\% at large depths
(8-10 km w.e.).

For depth -- intensity relation the bin width of 200 m w.e. has been
chosen. For depths more than 9 km w.e. the bin width was increased to
500 m w.e. to increase the statistics for each bin. The conversion of
muon intensity to the middle points of each depth bin has been done
using the formula:

\begin{equation}
I_{\mu}^m(x_i)=I_{\mu}^m(x_m) {{I_{\mu}^c(x_i)}\over{I_{\mu}^c(x_m)}}
\end{equation}

where $I_{\mu}^m(x_m)$ and $I_{\mu}^c(x_m)$ are the measured and calculated
muon intensities at the weighted average depth $x_m$ which 
corresponds to the depth bin with the middle value of $x_i$;
$I_{\mu}^m(x_i)$ and $I_{\mu}^c(x_i)$ are the derived and calculated
muon intensities at the depth $x_i$ which is the middle point of the
depth bin. The values of $x_m$ have been obtained by averaging the
depths for all angular bins contributing to the given depth bin
with a weight equal to the detected number of muons. To calculate
the muon intensities at $x_m$ and $x_i$ we have used again
eq. (\ref{Gaisser spectrum}) with the previously estimated parameters
of muon spectrum \cite{LVD,LVDR} and the simulated muon
survival probabilities.

Since the width of depth bins is quite small (200 m w.e. for depth bins
with high statistics) and the number of angular bins
contributing to each depth bin is quite large (several hundreds), 
the conversion factor does not exceed 10\%.

\section {`Depth -- vertical intensity' relation in Gran Sasso
rock}

`Depth -- vertical muon intensity' relation derived as it is described
in the previous section has been fitted with the calculated
distributions with 2 free parameters of the muon spectrum at sea level:
normalization constant, $A$, and the power index of the primary 
all-nucleon spectrum, $\gamma$. To calculate the muon intensities
underground we have used the formula:

\begin{equation}
I_{\mu}(x,\cos\theta)=\int_0^{\infty} P(E_{\mu 0},x)
{{d I_{\mu 0}(E_{\mu 0},\cos\theta)} \over {d E_{\mu 0}}}
d E_{\mu 0} \label{muon intensity}
\end{equation}

where $P(E_{\mu 0},x)$ is the survival probability of muon with an
initial energy $E_{\mu 0}$ at sea level to reach the depth $x$;
and ${{d I_{\mu 0}(E_{\mu 0},\cos\theta)} \over {d E_{\mu 0}}}$
is the muon spectrum at sea level at zenith angle
$\theta$. This spectrum has been taken according to 
\cite{Gaisser}:

\begin{eqnarray}
{{d I_{\mu 0} (E_{\mu 0}, \cos\theta)}\over{d E_{\mu 0}}}
& = &
A \cdot 0.14 \cdot E_{\mu 0}^{-\gamma} \nonumber \\
& \times &
\left({{1}\over{1+{{1.1 E_{\mu 0} \cos\theta^{\star}}\over{115 GeV}}}}+
{{0.054}\over{1+{{1.1 E_{\mu 0} \cos\theta^{\star}}\over{850 GeV}}}}
\right)
\label{Gaisser spectrum}
\end{eqnarray}

where the values of $cos\theta$ have been substituted by
$cos\theta^{\star}$ which have been taken from \cite{Volkovacos}.
In \cite{Volkovacos} $cos\theta^{\star}=
E_{\pi,K}^{cr}(cos\theta=1)/E_{\pi,K}^{cr}(cos\theta)$, where
$E_{\pi,K}^{cr}$ are the critical energies of pions and kaons.
This formula has been obtained under a simple assumption of
scaling in the high-energy hadron-nucleus interactions.
Under this assumption the power index of primary spectrum,
$\gamma$, is equal to that of meson (pion + kaon) spectrum,
$\gamma_{\pi,K}$. To fit the `depth -- vertical intensity'
relation measured by LVD we have put $\cos\theta=\cos\theta^{\star}=1$,
however, to convert the muon intensity to vertical (as it was
described in the previous section) we have used the values of
$cos\theta^{\star}$ from \cite{Volkovacos}.

We have used the muon survival probabilities $P(E_{\mu 0},x)$
calculated with the muon cross-sections from \cite{BBn,KP,KKP} and
with the account of stochasticity of all processes of muon
interaction with matter. They differ from those presented in \cite{LVDR}
and used also in \cite{LVD} (see also references therein) because
of the new corrections for muon bremsstrahlung and knock-on
electron production cross-sections proposed in \cite{KKP}
(see also the discussion in \cite{MUSIC}).

The measured `depth -- intensity' curve has been fitted with the
calculated function (see eq. (\ref{muon intensity})) with two
free parameters: additional normalization constant, $A$, and
the power index of primary all-nucleon spectrum, $\gamma$.
As a result of the fitting procedure the following values
of the free parameters have been obtained:
$A=1.95 \pm 0.31$, $\gamma=2.78 \pm 0.02$. These values are
in good agreement with the results of the analysis of
the depth -- angular distributions \cite{LVDP,propro}.
We note that the energy in eq. (\ref{Gaisser spectrum}) is expressed
in GeV and the intensity is expressed in cm$^{-2}$ s$^{-1}$ sr$^{-1}$.
The errors of the parameters include both
statistical and systematic uncertainties. The latter one takes into account
possible uncertainties in the depth, rock composition, density etc.,
but does not take into account the uncertainty in the cross-sections used
to simulate the muon transport through the rock. If we add the
uncertainty in the muon interaction cross-sections, the error of $\gamma$
will increase from 0.02 to 0.05 
and the error of $A$ to 1.0 (for the discussion about the uncertainty
due to different cross-sections see \cite{MUSIC}).
Similar analysis performed for single muons reveals
almost the same value of power index 
while the absolute intensity is 10\% smaller:
$A=1.65 \pm 0.30$, $\gamma=2.77 \pm 0.02$. 
We note that the estimates of the parameters $A$ and $\gamma$ are strongly
correlated. The larger the value of $\gamma$ is, the larger
the normalization factor $A$ should be.
The `depth -- vertical muon intesity' relation is shown in Figure 1
for all muon sample together with the best fit.
The muon intensities are presented also in Table ~\ref{fi:inttab} (column 2).

If the formula from \cite{Volkovasp} is used for the muon
spectrum at sea level instead of eq. (\ref{Gaisser spectrum}), the
best fit values of $\gamma$ will be decreased by 0.04-0.05 and will be
in agreement with the previously published values for single muons
\cite{LVD,LVDR}
analysed using the formula from \cite{Volkovasp}.

The value of $\gamma$ obtained with
LVD data is in reasonable agreement with the results of many other
surface and underground experiments (see, for example,
\cite{DEIS,MUTRON,ASD,MIPhI,MSU,NUSEX,MACROc}).

\section {`Depth -- vertical muon intensity' relation in standard
rock}

\indent The simulations carried out for Gran Sasso and standard rocks
allow us
to obtain the formula for the conversion of the depth in Gran Sasso
rock,
$x_{gs}$, to that in standard rock, $x_{st}$. This was done by
comparing the
values of $x_{st}$ and $x_{gs}$ for the same muon intensity:
$I_{\mu}(x_{st})=I_{\mu}(x_{gs})$. The muon intensities have been
calculated with the value of $\gamma$ which fit well the LVD data. 
The depth in standard rock can be
evaluated from the depth in Gran Sasso rock using the formula:

\begin{equation}
x_{st}=-9.344 + 1.0063 x_{gs} + 1.7835 \cdot 10^{-6} x_{gs}^2 -
5.7146 \cdot 10^{-11} x_{gs}^3,        \label{rock conversion}
\end{equation}

where the depth is measured in hg/cm$^2$. This formula is valid for
depth range 1--12 km w.e.. It has been used to convert the depth-- muon
intensity relation measured in Gran Sasso rock to that in standard 
rock to allow the comparison with the data of other experiments.

The `depth -- vertical intensity' relation in the standard rock
for all muon sample is presented in Figure 2.
It can be fitted with a three parameter function:

\begin{equation}
I_{\mu}(x)=A\left({{x_0}\over{x}}\right)^{\alpha}
\exp^{-{{x}\over{x_0}}},
\label{fit}
\end{equation}

where $A=(2.15 \pm 0.08) \cdot 10^{-6}$ cm$^{-2}$ s$^{-1}$ sr$^{-1}$,
$x_0=(1155^{+60}_{-30})$ hg/cm$^2$, $\alpha=1.93^{+0.20}_{-0.12}$.

The best fit function is also shown in Figure 2 by solid curve.
The LVD data converted to the standard rock agree quite well with
the best fit functions for the data of MACRO \cite{MACROc} 
(dashed curve in Figure 2) and Frejus \cite{Frejusf} (dotted curve
which almost coincide with the solid curve) underground experiments,
despite the difference in the formulae used for depth conversion.
The LVD data do not contradict also to the function which fit the data 
of NUSEX experiment \cite{NUSEX} (dash-dotted curve).

The muon intensities measured by LVD and converted to the standard rock
are presented also in Table ~\ref{fi:inttab} (3rd column).

\section {Muon energy spectrum at the sea level}

Using eq. (\ref{Gaisser spectrum}) with the estimates of the free
parameters, the LVD data for Gran Sasso rock can be converted to the
vertical muon spectrum at the sea level. The simple normalization
procedure has been applied:

\begin{equation}
I_{\mu 0}^m(x=0,E_{m0})={{I_{\mu}^m(x) I_{\mu 0}^c(x=0,E_{m0})}
\over{I_{\mu}^c(x)}},      \label{conversion to sea level}
\end{equation}

where $I_{\mu}^m(x)$ and $I_{\mu}^c(x)$ are the measured and calculated
vertical muon intensities at the depth $x$ in the Gran Sasso rock,
$I_{\mu 0}^m(x=0,E_{m0})$ and $I_{\mu 0}^c(x=0,E_{m0})$ are the
derived and calculated (using eq. (\ref{Gaisser spectrum})) 
differential vertical muon intensities at energy $E_{m0}$
at sea level. The median energies $E_{m0}(x)$ which determine
the muon intensity at the depth $x$ have been calculated using
the equation:

\begin{equation}
{{I_{\mu}(x)} \over {2}}=\int_0^{E_{m0}} P(E_{\mu 0},x) \cdot
{{d I_{\mu 0}(E_{\mu 0})} \over {d E_{\mu 0}}} \cdot
d E_{\mu 0}
\label{median energy}
\end{equation}

where ${{d I_{\mu 0}(E_{\mu 0})} \over {d E_{\mu 0}}}$ is
the muon energy spectrum at sea level.
The values of $I_{\mu 0}^m(x=0,E_{m0})$
derived from LVD data are presented in Table ~\ref{fi:spetab}.
The derived (full circles) and calculated (middle solid curve)
muon spectra
at the sea level are shown in Figure 3 together with the data of
MSU \cite{MSU} (diamonds), ASD \cite{ASD} (open circles) and
the best fit of MACRO \cite{MACROc} (dashed line). Upper
and lower solid curves in Figure 3 represent the errors in the parameters
and additional $10\%$ error in the absolute normalization of the
muon flux. Muon intensities at sea level derived from LVD data
are presented also in Table \ref{fi:spetab}.

\section {Conclusions}

The angular distribution of muon intensity measured by LVD has
been converted to the `depth -- vertical muon intensity' relation.
The analysis of this relation 
in the depth range 3000-12000 hg/cm$^2$ has been done
and the parameters of the muon spectrum at sea level have been
obtained: $A=1.9 \pm 1.0$, $\gamma=2.78 \pm 0.05$.
The errors include both statistical and systematic errors
with the systematic error due to the uncertainty of the muon
interaction cross-sections dominating.
Similar analysis performed for single muon events revealed almost
the same value of power index, while
the absolute intensity is 10\% smaller.
The `depth -- intensity' relation has been converted to the 
standard rock and fitted with a three-parameter function.
This relation agrees well with the data of other underground
experiments.
Using the measured `depth -- intensity' curve and the estimated
parameters of the muon spectrum at sea level we have derived the
vertical muon energy spectrum at sea level.

\section {Acknowledgements}

We wish to thank the staff of the Gran Sasso Laboratory
for their aid and collaboration. This work is supported by the
Italian Institute for Nuclear Phy\-sics (INFN) and in part by the
Italian Ministry of University and Scientific-Technological
Research (MURST), the Russian Ministry of Science and Technologies,
the Russian Foundation of Basic Research
(grant 96-02-19007), the US Department of Energy, the US National
Science Foundation, the State of Texas under its TATRP program,
and Brown University.

\pagebreak

\begin{figure}[htb]
\begin{center}
\epsfig{figure=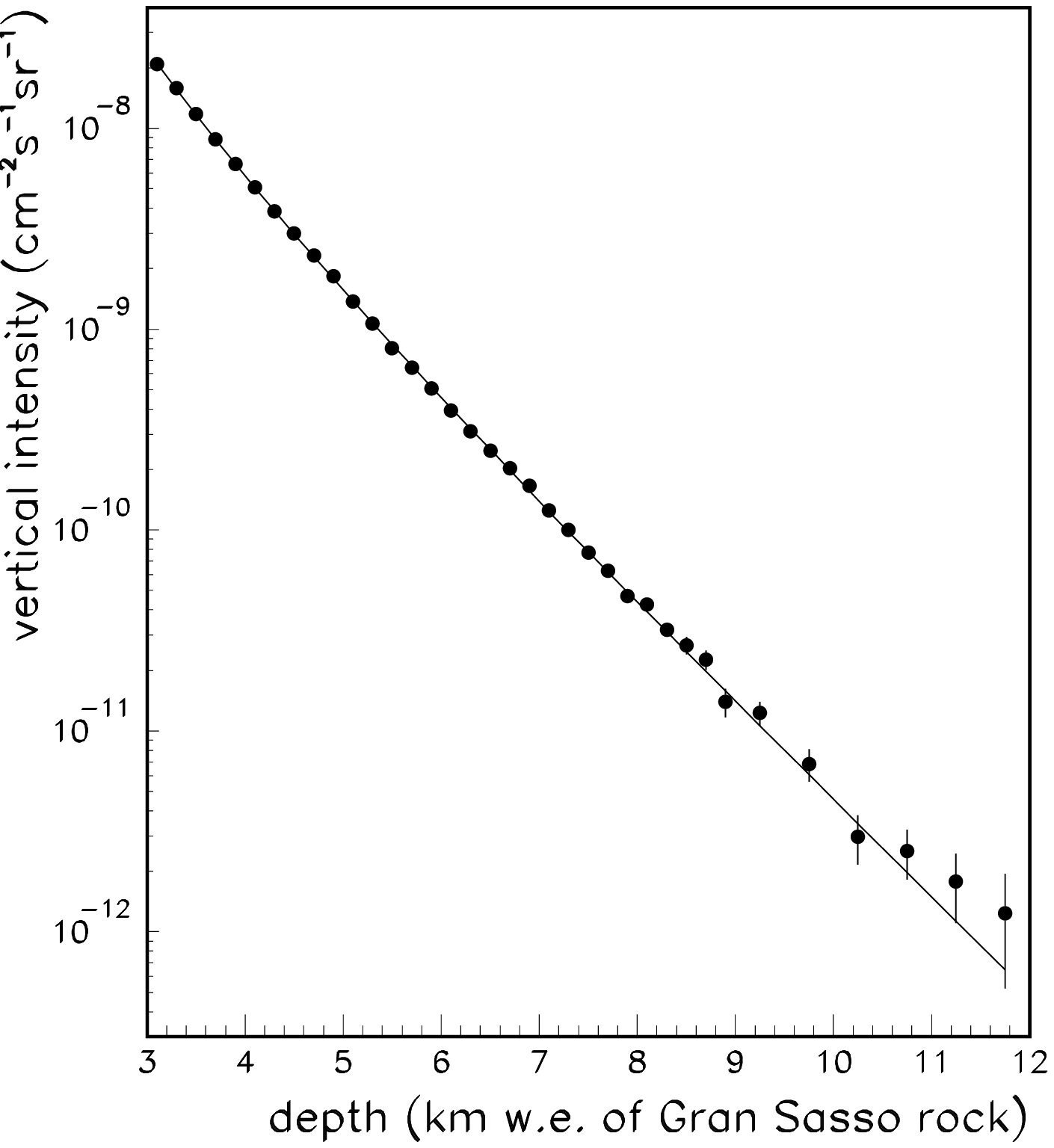,height=15cm}
\caption{ `Depth -- vertical muon intensity' curve in Gran Sasso rock
measured by LVD together with the best fit using 
eq. (\ref{Gaisser spectrum}) with the parameters:
$\gamma=2.78$ and $A=1.95$.}
\end{center}
\end{figure}

\pagebreak

\begin{figure}[htb]
\begin{center}
\epsfig{figure=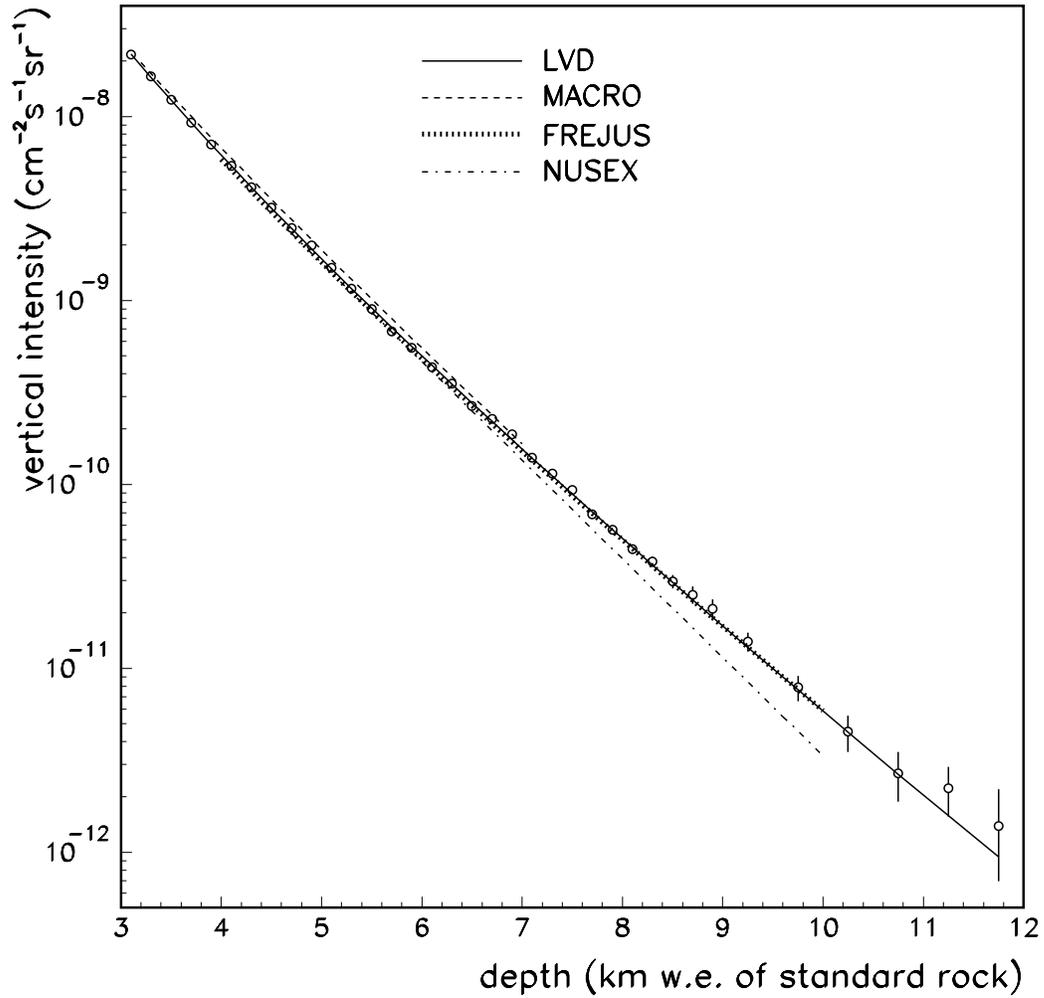,height=15cm}
\caption{ `Depth -- vertical muon intensity' curve in standard rock.
LVD data are presented together with the best fit using three-parameter
function (see eq. (\ref{fit})) and the best fits to the data
of other experiments: MACRO [11] (dashed curve),
Frejus [33] (dotted curve) and NUSEX [10] (dash-dotted
curve).}
\end{center}
\end{figure}

\pagebreak

\begin{figure}[htb]
\begin{center}
\epsfig{figure=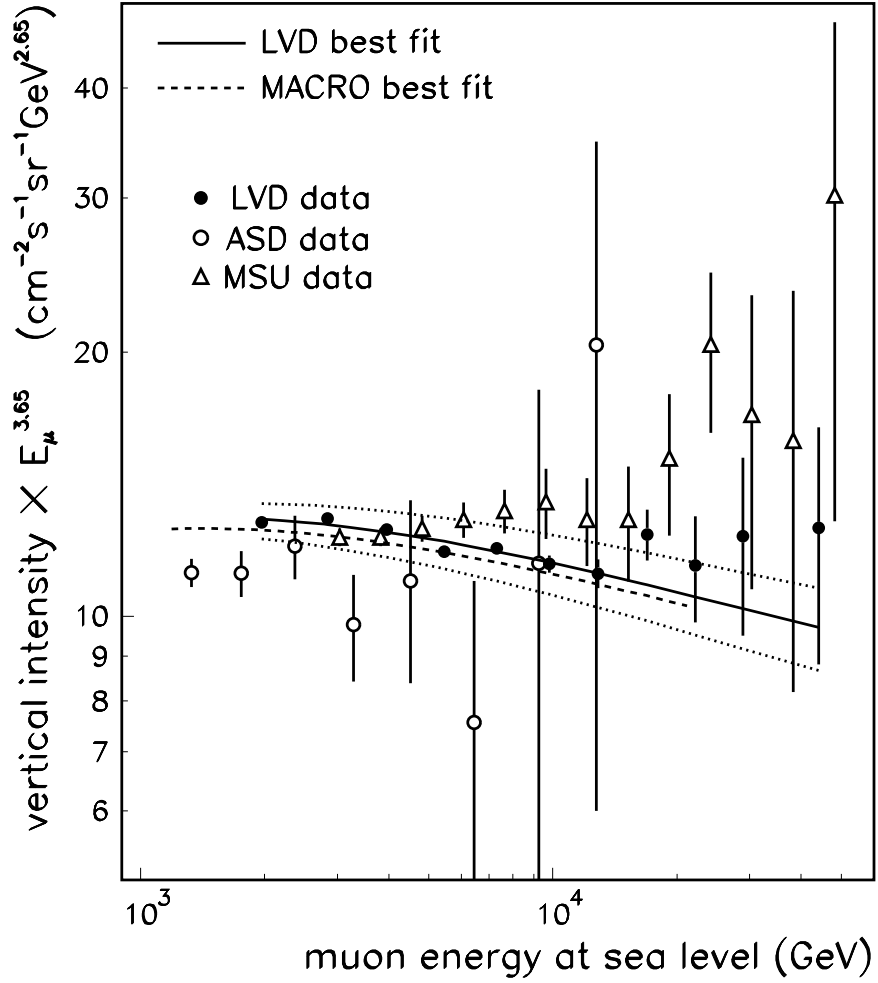,height=15cm}
\caption{Vertical muon energy spectra at sea level:
$\bullet$ -- LVD data; $\circ$ -- ASD data [4]; 
$\diamond$ -- MSU data [7]; solid curves --
LVD best fit together with the curves representing the errors
of the parameters and of the absolute flux normalization;
dashed curve -- MACRO best fit [11].}
\end{center}
\end{figure}

\pagebreak

\begin{table}
\caption{Vertical muon intensities measured by LVD vs depth
in the Gran Sasso and standard rocks. Errors include both
statistical and systematic uncertainties.}
\label{fi:inttab}
\begin{tabular}{|c|c|c|}   
$x$, $km.w.e.$ & $I_\mu$, $(cm^{2} s sr)^{-1}$, G.S. rock &
$I_\mu$, $(cm^{2} s sr)^{-1}$, st. rock \\
\hline
3.1 & $(2.09 \pm 0.02) \cdot 10^{-8}$ & $(2.17 \pm 0.02) \cdot 10^{-8}$ \\
3.3 & $(1.58 \pm 0.02) \cdot 10^{-8}$ & $(1.65 \pm 0.02) \cdot 10^{-8}$ \\
3.5 & $(1.18 \pm 0.01) \cdot 10^{-8}$ & $(1.23 \pm 0.01) \cdot 10^{-8}$ \\
3.7 & $(8.82 \pm 0.09) \cdot 10^{-9}$ & $(9.24 \pm 0.09) \cdot 10^{-9}$ \\
3.9 & $(6.66 \pm 0.07) \cdot 10^{-9}$ & $(7.04 \pm 0.07) \cdot 10^{-9}$ \\
4.1 & $(5.08 \pm 0.06) \cdot 10^{-9}$ & $(5.37 \pm 0.06) \cdot 10^{-9}$ \\
4.3 & $(3.87 \pm 0.04) \cdot 10^{-9}$ & $(4.12 \pm 0.04) \cdot 10^{-9}$ \\
4.5 & $(3.00 \pm 0.03) \cdot 10^{-9}$ & $(3.19 \pm 0.04) \cdot 10^{-9}$ \\
4.7 & $(2.34 \pm 0.03) \cdot 10^{-9}$ & $(2.47 \pm 0.03) \cdot 10^{-9}$ \\
4.9 & $(1.83 \pm 0.02) \cdot 10^{-9}$ & $(1.99 \pm 0.03) \cdot 10^{-9}$ \\
5.1 & $(1.37 \pm 0.02) \cdot 10^{-9}$ & $(1.50 \pm 0.02) \cdot 10^{-9}$ \\
5.3 & $(1.07 \pm 0.02) \cdot 10^{-9}$ & $(1.16 \pm 0.02) \cdot 10^{-9}$ \\
5.5 & $(8.07 \pm 0.13) \cdot 10^{-10}$ & $(8.99 \pm 0.14) \cdot 10^{-10}$ \\
5.7 & $(6.42 \pm 0.11) \cdot 10^{-10}$ & $(6.79 \pm 0.12) \cdot 10^{-10}$ \\
5.9 & $(5.07 \pm 0.09) \cdot 10^{-10}$ & $(5.52 \pm 0.10) \cdot 10^{-10}$ \\
6.1 & $(3.94 \pm 0.08) \cdot 10^{-10}$ & $(4.34 \pm 0.09) \cdot 10^{-10}$ \\
6.3 & $(3.11 \pm 0.07) \cdot 10^{-10}$ & $(3.53 \pm 0.08) \cdot 10^{-10}$ \\
6.5 & $(2.48 \pm 0.06) \cdot 10^{-10}$ & $(2.68 \pm 0.07) \cdot 10^{-10}$ \\
6.7 & $(2.03 \pm 0.07) \cdot 10^{-10}$ & $(2.27 \pm 0.07) \cdot 10^{-10}$ \\
6.9 & $(1.66 \pm 0.06) \cdot 10^{-10}$ & $(1.88 \pm 0.07) \cdot 10^{-10}$ \\
7.1 & $(1.25 \pm 0.05) \cdot 10^{-10}$ & $(1.40 \pm 0.05) \cdot 10^{-10}$ \\
7.3 & $(1.00 \pm 0.04) \cdot 10^{-10}$ & $(1.14 \pm 0.05) \cdot 10^{-10}$ \\
7.5 & $(7.70 \pm 0.34) \cdot 10^{-11}$ & $(9.32 \pm 0.39) \cdot 10^{-11}$ \\
7.7 & $(6.27 \pm 0.30) \cdot 10^{-11}$ & $(6.85 \pm 0.32) \cdot 10^{-11}$ \\
7.9 & $(4.68 \pm 0.25) \cdot 10^{-11}$ & $(5.65 \pm 0.29) \cdot 10^{-11}$ \\
8.1 & $(4.27 \pm 0.25) \cdot 10^{-11}$ & $(4.45 \pm 0.24) \cdot 10^{-11}$ \\
8.3 & $(3.19 \pm 0.25) \cdot 10^{-11}$ & $(3.80 \pm 0.26) \cdot 10^{-11}$ \\
8.5 & $(2.67 \pm 0.26) \cdot 10^{-11}$ & $(2.97 \pm 0.25) \cdot 10^{-11}$ \\
8.7 & $(2.26 \pm 0.26) \cdot 10^{-11}$ & $(2.51 \pm 0.27) \cdot 10^{-11}$ \\
8.9 & $(1.40 \pm 0.22) \cdot 10^{-11}$ & $(2.11 \pm 0.26) \cdot 10^{-11}$ \\
9.25 & $(1.23 \pm 0.17) \cdot 10^{-11}$ & $(1.40 \pm 0.16) \cdot 10^{-11}$ \\
9.75 & $(6.8 \pm 1.3) \cdot 10^{-12}$ & $(7.9 \pm 1.2) \cdot 10^{-12}$ \\
10.25 & $(3.0 \pm 0.8) \cdot 10^{-12}$ & $(4.5 \pm 1.0) \cdot 10^{-12}$ \\
10.75 & $(2.5 \pm 0.7) \cdot 10^{-12}$ & $(2.7 \pm 0.8) \cdot 10^{-12}$ \\
11.25 & $(1.8 \pm 0.7) \cdot 10^{-12}$ & $(2.2 \pm 0.7) \cdot 10^{-12}$ \\
11.75 & $(1.2 \pm 0.7) \cdot 10^{-12}$ & $(1.4 \pm 0.8) \cdot 10^{-12}$ \\
\end{tabular}  
\end{table}

\pagebreak

\begin{table}
\caption{Vertical muon spectrum at sea level derived from
LVD data. Errors include both statistical and systematic
uncertainties.}
\label{fi:spetab}
\begin{tabular}{|c|c|}
$E_{\mu 0}$, GeV & $I_{\mu}$, cm$^{-2}$ s$^{-1}$ sr$^{-1}$ GeV$^{-1}$\\
\hline
$1932$ & $(1.30 \pm 0.01) \cdot 10^{-11}$ \\
$2793$ & $(3.42 \pm 0.03) \cdot 10^{-12}$ \\
$3909$ & $(9.72 \pm 0.11) \cdot 10^{-13}$ \\
$5371$ & $(2.88 \pm 0.04) \cdot 10^{-13}$ \\
$7246$ & $(9.73 \pm 0.16) \cdot 10^{-14}$ \\
$9685$ & $(3.24 \pm 0.07) \cdot 10^{-14}$ \\
$12770$ & $(1.15 \pm 0.04) \cdot 10^{-14}$ \\
$16750$ & $(4.74 \pm 0.32) \cdot 10^{-15}$ \\
$21980$ & $(1.62 \pm 0.22) \cdot 10^{-15}$ \\
$28580$ & $(6.7 \pm 1.5) \cdot 10^{-16}$ \\
$42660$ & $(1.6 \pm 0.5) \cdot 10^{-16}$ \\
\end{tabular}
\end{table}


\begin{references}

\bibitem{MARS}
M.G.Thompson et al.,
{\em Proc. 15th Intern. Cosmic Ray Conf.\/} (Plovdiv) {\bf 6} (1977) 21.

\bibitem{DEIS}
O.C.Allkofer et al.,
{\em Proc. 17th Intern. Cosmic Ray Conf.\/} (Paris) {\bf 10} (1981) 321.

\bibitem{MUTRON}
S.Matsuno et al.,
{\em Phys. Rev. D\/} {\bf 29} (1984) 1.

\bibitem{ASD}
F.F.Khalchukov et al.,
{\em Proc. 19th Intern. Cosmic Ray Conf.\/} (La Jolla) {\bf 8} (1985)
12;
R.I.Enikeev et al.,
{\em Sov. J. Nucl. Phys.\/} {\bf 47} (1988) 1044.

\bibitem{Baksans}
V.N.Bakatanov et al.,
{\em Proc. 21st Intern. Cosmic Ray Conf.\/} (Adelaide) {\bf 9} (1990)
375.

\bibitem{MIPhI}
V.D.Ashitkov et al.,
{\em Proc. 19th Intern. Cosmic Ray Conf.\/} (La Jolla) {\bf 8} (1985)
77.

\bibitem{MSU}
N.P.Il'ina et al.,
{\em Proc. 24st Intern. Cosmic Ray Conf.\/} (Rome) {\bf 1}
(1995) 524.

\bibitem{KGF}
M.R.Krishnaswami et al.,
{\em Proc. 18th Intern. Cosmic Ray Conf.\/} (Bangalore) {\bf 11}
(1983) 450.

\bibitem{Baksan}
Yu.M. Andreyev, V.I.Gurentsov, and I.M.Kogai,
{\em Proc. 20th Intern. Cosmic Ray Conf.\/} (Moscow) {\bf 6} (1987) 200.

\bibitem{NUSEX}
G.Battistoni et al.,
{\em Nuovo Cimento\/} {\bf 9C} (1986) 196.

\bibitem{MACROc}
M.Ambrosio et al. (MACRO Collaboration),
{\em Phys. Rev. D\/} {\bf 52} (1995) 3793.

\bibitem{LVD}
M.Aglietta et al. (LVD Collaboration),
{\em Astroparticle Phys.\/} {\bf 3} (1995) 311.

\bibitem{Gurentsova}
V.I.Gurentsov, G.T.Zatsepin, and E.D.Mikhalchi,
{\em Sov. J. Nucl. Phys.\/} {\bf 23} (1976) 1001.

\bibitem{Gurentsov}
V.I.Gurentsov,
{\em Preprint INR, P-0380\/} (1984) (in Russian).

\bibitem{Castellina}
H.Bilokon et al.,
{\em Nucl. Instrum. and Meth. in Phys. Res.\/} {\bf A303} (1991) 381.

\bibitem{Bilokon}
H.Bilokon et al.,
{\em Preprint LNGS 94/92\/} (1994).

\bibitem{Lipari}
P.Lipari and T.Stanev,
{\em Phys. Rev. D\/} {\bf 44} (1991) 3543.

\bibitem{Kudryavtsev}
V.A.Kudryavtsev,
{\em Preprint INR, P-0529\/} (1987) (in Russian).

\bibitem{Naumov}
E.V.Bugaev et al.,
{\em Preprint INR, P-0347\/} (1984) (in Russian).

\bibitem{Lagutin}
A.A.Lagutin,
{\em Preprint ASU, 94/1\/} (1994).

\bibitem{LVDR}
M.Aglietta et al. (LVD Collaboration),
{\em Proc. 24th Intern. Cosmic Ray Conf.\/} (Rome) {\bf 1} (1995) 557.

\bibitem{BBb}
L.B.Bezrukov and E.V.Bugaev,
{\em Proc. 17th Intern. Cosmic Ray Conf.\/} (Paris) {\bf 7} (1981) 102.

\bibitem{BBn}
L.B.Bezrukov and E.V.Bugaev,
{\em Proc. 17th Intern. Cosmic Ray Conf.\/} (Paris) {\bf 7} (1981) 90.

\bibitem{KP}
R.P.Kokoulin and A.A.Petrukhin,
{\em Proc. 12th Intern. Cosmic Ray Conf.\/} (Hobart) {\bf 6} (1971)
2436.

\bibitem{KKP}
S.R.Kelner, R.P.Kokoulin, and A.A.Petrukhin,
{\em Physics of Atomic Nuclei\/} {\bf 60} (1997) 576.

\bibitem{LVDP}
LVD Collaboration (M.Aglietta et al.),
{\em Proc. 25th ICRC\/} (Durban) {\bf 6} (1997) 341.

\bibitem{propro}
LVD Collaboration (M.Aglietta et al.),
to be submitted to {\em Phys. Rev. D}. 

\bibitem{LVDa}
M.Aglietta et al.,
{\em Astroparticle Phys.\/} {\bf 2} (1994) 103.

\bibitem{Gaisser}
T.K.Gaisser,
{\em Cosmic Rays and Particle Physics\/}
(Cambridge University Press, 1990)

\bibitem{Volkovacos}
L.V.Volkova,
{\em Preprint Lebedev Physical Institute N 72\/} (1969).

\bibitem{MUSIC}
P.Antonioli et al.,
{\em Astroparticle Phys.\/} {\bf 7} (1997) 357.

\bibitem{Volkovasp}
L.V.Volkova, G.T.Zatsepin, and L.A.Kuzmichev,
{\em Sov. J. Nucl. Phys.\/} {\bf 29} (1979) 1252.

\bibitem{Frejusf}
Ch.Berger et al.,
{\em Phys. Rev. D\/} {\bf 40} (1989) 2163.

\end{references}
\end{document}